# The ratio of top scientists to the academic staff as an indicator of the competitive strength of universities[1]

*Giovanni Abramo*[*]

Laboratory for Studies of Research and Technology Transfer, Institute for System Analysis and Computer Science (IASI-CNR), National Research Council of Italy,
Viale Manzoni 30, 00185 Rome, Italy
giovanni.abramo@uniroma2.it

*Ciriaco Andrea D'Angelo*

Department of Engineering and Management, University of Rome 'Tor Vergata' and Laboratory for Studies of Research and Technology Transfer, Institute for System Analysis and Computer Science (IASI-CNR)
Via del Politecnico 1, 00133 Rome, Italy
dangelo@dii.uniroma2.it

*Anastasiia Soldatenkova*

Department of Engineering and Management, University of Rome 'Tor Vergata'
Via del Politecnico 1, 00133 Rome, Italy
anastasiia.soldatenkova@uniroma2.it

**Abstract**

The ability to attract and retain talented professors is a distinctive competence of world-class universities and a source of competitive advantage. The ratio of top scientists to academic staff could therefore be an indicator of the competitive strength of the universities. This work identifies the Italian top scientists in over 200 fields, by their research productivity. It then ranks the relative universities by the ratio of top scientists to overall faculty. Finally, it contrasts this list with the ranking list by average productivity of the overall faculty. The analysis is carried out at the field, discipline, and overall university levels. The paper also explores the secondary question of whether the ratio of top scientists to faculty is related to the size of the university.

**Keywords**

*Research evaluation; bibliometrics; rankings; FSS; Italy*

[*] *Corresponding author*

## 1. Introduction

The single most important distinctive competence of world-class universities is probably their ability to attract and retain highly qualified faculty (Salmi, 2009; Mohrman et al., 2008). Talented professors attract talented students. They produce groundbreaking research results, which in turn attract donations, public and private research funds, venture capital and establishment of high-tech companies in the territory. The outstanding reputation, abundance of financial resources, and highly attractive research environment feed a virtuous circle, leading to a sustained competitive advantage over other universities. Competitive higher education systems, such as those observed in English speaking nations, are therefore likely to present a number of universities with distinctive achievement in quality of education and research. This supposition is supported by a glance at the most popular world university "league tables", such as the Academic Ranking of World Universities (ARWU), the Times Higher Education (THE) World Academic Rankings, the CWTS Leiden Rankings, and the SCImago Institutions Rankings (SIR). Several studies call into question the methodology and the relevance of the indicators used in these rankings (Abramo et al., 2011; Butler, 2010; Dehon, 2010; Turner, 2005; van Raan, 2005). However, Saisana et al. (2011), while accepting that the ARWU and the THE should not be used to compare the performance of individual universities, demonstrate that their rankings are reliable for the top 10 universities. Indeed the top 10 institutions in both the 2015 ARWU[2] and THE 2015-2016 ranking[3] consist entirely of ones situated in the U.S. or U.K., and the same also occurs in the 2015 SIR and Leiden size-independent rankings.

The U.S. and U.K. have experienced an evolution of national policies that favor the birth and development of true markets in higher education. In contrast, many European nations witness an excess of public control, inhibiting the initiation of competitive mechanisms and leading to the development of generally undifferentiated higher education systems that are unable to compete at a global level for access to economic and human resources (public and private funds, talented students, excellent faculty) (Veugelers and van der Ploeg, 2008). Auranen and Nieminen (2010) report that in countries such as Germany, Sweden and Denmark, there is still no expectation of distinguishing different levels of excellence among universities, due to the almost total lack of competition among the actors in the systems. (The present authors observe the same situation in Italy.)

Recently, scientometricians have inquired into the distribution of talented scientists among universities. Because top-level scientists contribute more than unproductive ones to the overall research performance of universities (Abramo et al., 2013a), logic would have it that in competitive systems, top research universities would achieve this status from a concentration of top scientists (TSs). However, what is the case in education systems with little differentiation? Are the TSs quite uniformly distributed among universities? Common sense would lead us to expect this, otherwise the universities with a high concentration of TSs would have to experience an even higher concentration of unproductive scientists in order to remain undifferentiated, a fact which seems difficult to accept. The intention of the current work is to answer this question of the distribution of top scientists in undifferentiated higher education systems, taking the

---

[2] http://www.shanghairanking.com/ARWU2015.html, last accessed on April 19, 2016.
[3] https://www.timeshighereducation.com/world-university-rankings/2016/world-ranking#!/page/0/length/25, last accessed on April 19, 2016.

Italian case as specific reference. Italy offers a classic example of an undifferentiated higher education system, as shown by Bonaccorsi and Cicero (2015), Abramo et al. (2012a), and confirmed through a further sophisticated assessment methodology, accounting for uncertainty in the measure of research performance (Abramo et al., 2015a).

The literature on the distribution of TSs among universities is quite scarce. To the best of our knowledge, the only work to specifically focus on this subject is by Yang et al. (2015). These authors identify the world's top scientists and institutions in twenty broad research fields, by total counts of citations for publications indexed in the Web of Science (WoS) in the period 2008-2011. Their findings are that in the larger fields, more than 80% of TSs work at top institutions, although the concentration is less for smaller fields such as mathematics and computer science. Other research, tangentially related to the topic, has been by Bornmann and Bauer (2015), who rank institutions by the total number of highly cited researchers, and by Abramo and D'Angelo (2015), who rank universities by the number of highly-cited articles per scientist.

The current study overcomes what in our view are two major limits of Yang et al. (2015). The first concerns the performance indicator used to identify the top research institutions. In fact Yang et al. adopt the measure of the total count of citations, but this is a size-dependent indicator, which inevitably favors large institutions. In this study we employ a size-independent indicator, specifically fractional scientific strength (FSS), which is a productivity indicator embedding both the fractional counting of publications and their field-normalized citations (Abramo and D'Angelo, 2014). The second limit concerns the results under examination. What interests Yang et al. is the whole numbers of TSs employed in the top institutions (or the percentages of top scientists out of the overall population). Once again, all others equal, the larger the size of the institution, the greater the chance of employing the higher number of TSs. We will instead calculate, for each university, the ratio of TSs to the overall faculty. This change provides an indicator of the competitive strength of the universities, and responds to our first research question.

Further, we will then investigate the specific aspect of the relation between the institution's average research performance and its ratio of TSs to overall academic staff. We will also verify the side question of whether there are varying returns to size. The analytical approach involves a fine-grained analysis in over 200 research fields, meaning that the analysis is ten times finer than that of Yang et al. (2015). Given that the intensity of publication varies across fields (Butler 2007; Abramo and D'Angelo, 2007; Moed et al., 1985; Garfield, 1979), this approach has the particular advantage of avoiding distortions due to the coarse aggregation of research fields (Abramo et al., 2008).

In the next sections of the paper we describe the distinctive features of the Italian higher education system and then present the data and methods used. The final sections provide the results of the analysis, leading to the conclusions.

## 2. The Italian higher education system

The Italian Ministry of Education, Universities, and Research (MIUR) recognizes a total of 96 universities as having the authority to issue legally recognized degrees. Of these, 29 are small, private, special-focus universities, of which 13 offer only e-learning, 67 are public and generally multi-disciplinary universities, scattered throughout Italy. Six of them are *Scuole Superiori* (Schools for Advanced Studies), specifically devoted to highly talented students, with very small faculties and tightly limited enrolment per degree program. In the overall system, 94.9% of faculty are employed in public universities (0.5% in *Scuole Superiori*). Public universities are largely financed by the government through non-competitive allocation of funds. Until 2009 the core government funding (56% of universities' total income) was input oriented (i.e. independent of merit, and distributed to universities in a manner intended to satisfy the needs of each and all equally, with respect to their size and research disciplines). It was only following the first national research evaluation exercise concluded in 2006, that a minimal share, equivalent to 3.9% of total income, was assigned by the MIUR as a function of the assessment of research and teaching.

Despite interventions intended to grant increased autonomy and responsibilities to the universities (Law 168 of 1989)[4], the Italian higher education system is a long-standing, classic example of a public and highly centralized governance structure, with low levels of autonomy at the university level and a very strong role played by the central state.

In keeping with the Humboldtian model, there are no 'teaching-only' universities in Italy, as all professors are required to carry out both research and teaching. National legislation includes a provision that each faculty member must provide a minimum of 350 hours per year of teaching. At the close of 2015, there were 54,800 faculty members in Italy (full, associate and assistant professors) and a roughly equal number of technical-administrative staff. All new personnel enter the university system through public competitions, and career advancement depends on further public competitions.

Salaries are regulated at the central level and are calculated according to role (administrative, technical or professorial), rank within role (e.g. assistant, associate or full professor) and seniority. None of a professor's salary depends on merit. Moreover, as in all Italian public administration, dismissal of unproductive professors is unheard of.

The entire legislative-administrative context has created a culture that is hardly competitive, yet flourishing with favoritism and other opportunistic behaviors that are dysfunctional to the social and economic roles of the academia (Zagaria, 2007; Perotti, 2008). Abramo et al. (2014) investigated 287 associate professor competitions. The analysis showed several critical issues, particularly concerning unsuccessful candidates who outperformed the competition winners in terms of research productivity, as well as a number of competition winners who resulted as totally unproductive. Almost half of individual competitions selected candidates who would go on to achieve below-median productivity in their field of research over the subsequent triennium. A more recent work (Abramo et al., 2015b) showed that the fundamental determinant of an academic candidate's success is not scientific merit, but rather the number of years that the candidate has belonged to the same university as the president of the selection committee. Thus, universities are unable to attract significant numbers of talented foreign faculty: only 1.8% of research staff are foreign nationals. Over the period 2009-2013, 3,178 (9.1%) of the 34,862 professors in the Sciences did not publish any scientific articles in WoS indexed journals. Another 868 professors (2.5%) achieved publication, but their work was never cited. This means that 4,046 individuals (11.6%) had no impact on scientific progress measurable by bibliometric databases.[2] This share of unproductive faculty has been declining but is still too high, particularly given that the legislative structure obligates all professors to conduct research. Indeed, highly competitive academic systems typically present instances of distinct "research" and "teaching" universities, but in Italy all universities are intended to serve both purposes, with all of them staffed by professors who are responsible for both research and teaching.

---

[4] http://www.normattiva.it/uri-res/N2Ls?urn:nir:stato:Legge:1989-05-09;168, last accessed on April 19, 2016.

## 3. Data and methods

The intention of our analysis is to provide and examine the ranking all Italian universities according to their ratio to total faculty of those professors that rank in the top 10% for productivity, out of all Italian professors in the same field.

We restrict the analysis to those fields where the prevalent form of codification for research output is publication in scientific journals, and therefore bibliometrics can be applied to measure research performance. For brevity, we call those fields the Sciences, and distinguish them from the Social Sciences and Arts & Humanities. In the Italian university system all professors are classified in one and only one field, named the scientific disciplinary sector (SDS), 370 in all. SDSs are grouped into disciplines, named university disciplinary areas (UDAs), 14 in all[5]. 192 such fields, grouped into nine UDAs[6], fall in the Sciences.

Data on the faculty at each university and their SDS classification were extracted from the database on Italian university personnel, maintained by the MIUR. The bibliometric dataset used to measure performance is extracted from the Italian Observatory of Public Research, a database developed and maintained by the present authors and derived under license from the Thomson Reuters WoS. Beginning from the raw data of the WoS, and applying a complex algorithm to reconcile the author's affiliation and disambiguate the true identity of the authors, each publication (article, article review and conference proceeding) is attributed to the university scientist or scientists that produced it (D'Angelo et al. 2011). Thanks to this algorithm, we can produce rankings of research performance at the individual level, on a national scale.

In this work we measure the research performance in the publication period 2009-2013. As said above, the indicator of performance that we use for professors and universities is $FSS$. At the professor level, we calculate $FSS_P$, as follows:

$$FSS_P = \frac{1}{t}\sum_{i=1}^{N}\frac{c_i}{\bar{c}}f_i$$

[1]

Where:
$t$ = number of years of work in the period under observation
$N$ = number of publications in the period under observation
$c_i$ = citations received by publication $i$ (counted at 31/05/2015)

---

[5] The complete list is accessible on http://attiministeriali.miur.it/UserFiles/115.htm, last accessed on April 19, 2016.
[6] Mathematics and computer sciences, Physics, Chemistry, Earth sciences, Biology, Medicine, Agricultural and veterinary sciences, Civil engineering, Industrial and information engineering

$\bar{c}$ = average of distribution of citations received for all cited publications[7] in same year and subject category of publication $i$

$f_i$ = fractional contribution of professor to publication $i$.

The fractional contribution equals the inverse of the number of authors in those fields where the practice is to place the authors in simple alphabetical order but assumes different weights in other cases. For the life sciences, widespread practice in Italy is for the authors to indicate the various contributions to the published research by the order of the names in the listing of the authors. So for the life science, we give different weights to each co-author according to their position in the list of authors and the character of the co-authorship (intra-mural or extra-mural) (Abramo et al. 2013b). If the first and last authors belong to the same university, 40% of the citation is attributed to each of them, the remaining 20% is divided among all other authors. If the first two and last two authors belong to different universities, 30% of the citation is attributed to the first and last authors, 15% of the citation is attributed to the second and last authors but one, the remaining 10% is divided among all others[8].

A thorough description of the economic theory underlying the operationalization of FSS, together with the assumptions and limits of the measurement, can be found in Abramo and D'Angelo (2014)[9].

Based on the value of *FSS* we obtain, for each SDS, a ranking list of all professors. We define TSs as those that place from the 90 percentile up.

To analyze the relation between the ratio of TSs to the faculty and the performance of the university, we need to assess the research performance of universities. To do that, we first normalize the FSS of each professor to the mean of all Italian productive professors in the same SDS, and then average the normalized FSS to overall faculty. In formulae, the productivity $FSS_U$ over a certain period for university $U$ is:

$$FSS_U = \frac{1}{RS}\sum_{j=1}^{RS}\frac{FSS_{R_j}}{\overline{FSS_R}}$$

[2]

Where:

$RS$ = number of professors of university $U$, in the observed period;

$FSS_{R_j}$ = productivity of professor $j$;

$\overline{FSS_R}$ = national average productivity of all productive professors in the same SDS of professor $j$.

## 4. Results and analysis

In this section we prepare and analyze the rankings of universities. After displaying our dataset, we calculate the ratio of TSs to faculty for each university, at the SDS and UDA levels. We then rank the Italian universities by TS ratio and analyze the distribution of the ratio, in response to our first research question, concerning the distribution of top scientists in undifferentiated higher education systems. We follow on to investigate whether the size of universities affects the value of the TS ratio (varying returns to size). Finally, we contrast the universities rankings by TS ratio and average research performance ($FSS_U$).

---

[7] Abramo et al. (2012b) demonstrated that the average of the distribution of citations received for all cited publications of the same year and subject category is the most effective scaling factor.

[8] The weightings were assigned following advice from senior Italian professors in the life sciences. The values could be changed to suit different practices in other national contexts.

[9] The reader may notice that, differently from the formula found in the referenced article, in this work we do not normalize the total impact by capital (salary of the professor). The reason is that we want to identify the top producers, regardless their cost.

Table 1 shows the dataset at the UDA level. For each UDA we report the number of universities with at least 10 faculty members, the total number of professors and the amount of TSs. We recall that TSs are defined here as those scientists above the 90[th] percentile in the ranking list of all Italian academic staff in the same SDS. The overall dataset concerns 64 universities and 34,862 professors, of which 3,571 fall in the category of TS.

*Table 1: Dataset for the analysis - number of universities, faculty and top scientists in each UDA*

| UDA | Universities | Faculty | Top scientists* |
|---|---|---|---|
| 1 - Mathematics and computer science | 50 | 3,268 | 332 |
| 2 - Physics | 43 | 2,333 | 237 |
| 3 - Chemistry | 42 | 2,996 | 303 |
| 4 - Earth sciences | 31 | 1,114 | 115 |
| 5 - Biology | 53 | 4,971 | 507 |
| 6 - Medicine | 42 | 10,370 | 1,062 |
| 7 - Agricultural and veterinary sciences | 29 | 3,076 | 319 |
| 8 - Civil engineering | 36 | 1,535 | 158 |
| 9 - Industrial and information engineering | 49 | 5,199 | 538 |
| Total | 64 | 34,862 | 3,571 |

* The number of top scientists does not equal exactly 10% of the research staff in each UDA because of ties in the ranking lists.

After identifying the TSs in each SDS, through their affiliation we are able to measure for every particular university the ratio of TSs to the overall faculty. As an example, in Table 2 we report the case of UNIV_45 in the SDSs of UDA 1 (Mathematics and computer science). About a quarter of the faculty are TSs, however the TS ratios in each SDS are uneven: there are no TSs in two out of six SDSs.

*Table 2: Ratio of top scientists to overall faculty in each SDS of UDA 1 for the case of UNIV_**45***

| SDS* | Research staff | TS ratio |
|---|---|---|
| MAT/02 - Algebra | 3 | 33.3% |
| MAT/03 - Geometry | 9 | 44.4% |
| MAT/05 - Mathematical analysis | 16 | 0% |
| MAT/07 - Mathematical physics | 3 | 0% |
| MAT/08 - Numerical analysis | 6 | 50.0% |
| INF/01 - Computer science | 11 | 36.4% |
| Total | 52 | 23.1% |

* The SDSs with less than 3 professors were excluded

Table 3 instead presents the TS ratios in each UDA and the corresponding national rank. In four UDAs, UNIV_45 employs less than 10 professors and the related statistics are not shown. The total number of TSs is 47, representing a TS ratio of 19.7% of overall faculty in the five UDAs analyzed. Univ_45 ranks in the national top 5% in UDA 1 (Mathematics and computer science), and UDA 9 (Industrial and information engineering), but below the median in UDA 5 (Biology). In the overall ranking by TS ratio, UNIV_45 ranks fifth out of 64 universities.

*Table 3: Ratio of top scientists to overall faculty in each UDA for the case of UNIV_45*

| UDA* | Research staff | TS ratio | National rank | Percentile |
|---|---|---|---|---|
| 1 | 52 | 23.1% | 3 out of 50 | 95.9% |
| 2 | 38 | 18.4% | 6 out of 43 | 88.1% |
| 5 | 14 | 7.1% | 36 out of 53 | 32.7% |
| 8 | 39 | 10.3% | 16 out of 36 | 57.1% |
| 9 | 75 | 25.3% | 3 out of 49 | 95.8% |
| All | 239 | 19.7% | 5 out of 64 | 93.7% |

*\* The UDAs with less than 10 professors were excluded. 1 = Mathematics and computer science; 2 = Physics; 5 = Biology; 8 = Civil engineering; 9 = Industrial and information engineering*

Figure 1 plots each university by its TS ratio and the corresponding position in the ranking list. With the help of the gaps in the density of the graph we can observe the differences in the TS ratios between neighboring universities in the rank. The first four universities in the ranking list stand out clearly. Then, starting from the 6th position, the TS ratio decreases quite evenly (average difference between neighboring positions is 0.2% TS ratio). The highest intervals appear between the first and the second, and the fourth and the fifth universities in the ranking list, with gaps in TS ratio of 12.5% and 8.5% respectively.

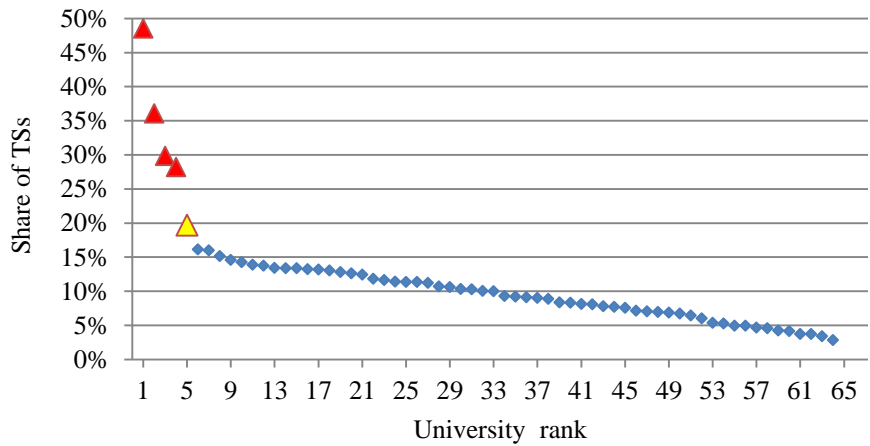

*Figure 1 Scatter plot of universities by ratio of top scientists to faculty and position in the ranking list; triangles indicate outliers*

As a side question, we investigate also whether the TS ratios vary more than proportionally with the size of the university. In a previous paper we showed no evidence of returns of productivity to size (Abramo et al., 2012c). We take the opportunity here to verify whether the same holds true for the TS ratio. Figure 2 shows a scatter plot of universities, positioned by TS ratio and the size of faculty. The Pearson correlation coefficient is equal to -0.129, demonstrating negligible linear relationship between the two variables. The four outliers by TS ratio are small-sized universities, each with less than 70 professors in the UDAs analyzed, while the median of the distribution is 420 and the maximum is 2642. Calculating the correlation coefficient without the top four universities by TS ratio and the largest one by size, the correlation remains weak ($\rho = 0.226$). We now contrast the universities ranking list by TS ratio with that by productivity ($FSS_U$). We first show the analysis at UDA level, and then at overall university level. Table 4 presents the comparison for Agricultural and veterinary sciences (UDA 7). Here, 55% of universities occupy the same positions in both rankings. Although the correlation between the two rankings is very strong (Spearman ρ = 0.861), there still occur noticeable shifts in rank for several universities (on average 13.2%). The maximum percentile shift equals 35.7%, corresponding to a two-quartile shift (UNIV_43).

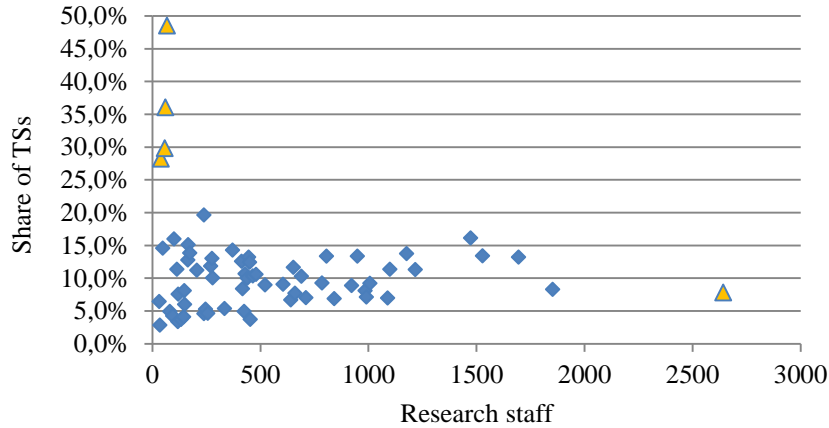

*Figure 2 Scatter plot of universities positioned by top scientists ratio and faculty size; triangles indicate outliers*

*Table 4: Comparison of university ranking lists by TS ratio and $FSS_U$ in UDA 7 (Agricultural and veterinary sciences)*

| University | Research staff | TS ratio | Rank | FSS | Rank | Rank Shift | Percentile shift | Quartile Shift |
|---|---|---|---|---|---|---|---|---|
| UNIV_26 | 64 | 25.0% | 1 | 1.175 | 3 | 2 | 7.1 | 0 |
| UNIV_8 | 12 | 25.0% | 1 | 1.113 | 4 | 3 | 10.7 | 0 |
| UNIV_18 | 93 | 17.2% | 3 | 1.306 | 2 | 1 | 3.6 | 0 |
| UNIV_2 | 12 | 16.7% | 4 | 1.016 | 6 | 2 | 7.1 | 0 |
| UNIV_12 | 72 | 16.7% | 4 | 0.942 | 13 | 9 | 32.1 | 1 |
| UNIV_34 | 184 | 16.3% | 6 | 1.321 | 1 | 5 | 17.9 | 0 |
| UNIV_44 | 189 | 13.8% | 7 | 0.981 | 10 | 3 | 10.7 | 1 |
| UNIV_49 | 22 | 13.6% | 8 | 0.956 | 11 | 3 | 10.7 | 1 |
| UNIV_39 | 15 | 13.3% | 9 | 0.843 | 15 | 3 | 21.4 | 0 |
| UNIV_56 | 255 | 13.3% | 9 | 0.956 | 12 | 6 | 10.7 | 0 |
| UNIV_29 | 273 | 13.2% | 11 | 0.991 | 8 | 3 | 10.7 | 1 |
| UNIV_28 | 64 | 12.5% | 12 | 1.024 | 5 | 7 | 25.0 | 1 |
| UNIV_32 | 209 | 12.4% | 13 | 0.988 | 9 | 4 | 14.3 | 0 |
| UNIV_62 | 45 | 11.1% | 14 | 0.734 | 17 | 3 | 10.7 | 1 |
| UNIV_20 | 195 | 10.8% | 15 | 1.004 | 7 | 8 | 28.6 | 1 |
| UNIV_60 | 139 | 8.6% | 16 | 0.817 | 16 | 0 | 0.0 | 0 |
| UNIV_35 | 108 | 8.3% | 17 | 0.687 | 20 | 3 | 10.7 | 0 |
| UNIV_58 | 105 | 7.6% | 18 | 0.695 | 19 | 1 | 3.6 | 0 |
| UNIV_47 | 103 | 6.8% | 19 | 0.731 | 18 | 1 | 3.6 | 0 |
| UNIV_17 | 74 | 6.8% | 20 | 0.593 | 24 | 4 | 14.3 | 1 |
| UNIV_36 | 75 | 6.7% | 21 | 0.561 | 25 | 4 | 14.3 | 1 |
| UNIV_25 | 146 | 5.5% | 22 | 0.530 | 27 | 5 | 17.9 | 1 |
| UNIV_38 | 153 | 4.6% | 23 | 0.658 | 22 | 1 | 3.6 | 1 |
| UNIV_43 | 73 | 4.1% | 24 | 0.891 | 14 | 10 | 35.7 | 2 |
| UNIV_41 | 135 | 3.7% | 25 | 0.600 | 23 | 2 | 7.1 | 0 |
| UNIV_13 | 55 | 3.6% | 26 | 0.487 | 28 | 2 | 7.1 | 0 |
| UNIV_52 | 64 | 3.1% | 27 | 0.441 | 29 | 2 | 7.1 | 0 |

| UNIV_57 | 38 | 2.6% | 28 | 0.659 | 21 | 7 | 25.0 | 1 |
| UNIV_31 | 18 | 0.0% | 29 | 0.532 | 26 | 3 | 10.7 | 0 |

Table 5 shows the ranking lists of Italian universities in all UDAs by TS ratio and by $FSS_U$. Four universities outperform the others by both TS ratio and $FSS_U$: they are a private university focused in medicine and three Schools for Advanced Studies. The correlation between them is very strong (Spearman $\rho = 0.924$), however there are numerous universities that shift rank (only 9 out of 64 universities occupy the same position in both rankings). The maximum shift is two quartiles, occurring in two cases, with the largest leaps equaling 42.8 (UNIV_43) and 38.1 (UNIV_2) percentiles, while the average is 7.8%. UNIV_43 presents a very low TS ratio ranking, 59 out of 64. This low ratio does not jeopardize its rank (32) by the average productivity of all its professors ($FSS_U$), revealing a low dispersion of performance among the faculty. The opposite is true for UNIV_2, whereby the high share of TSs (rank = 9) cannot make up for the comparably lower productivity of the rest of the faculty (overall rank by $FSS_U$ = 33), revealing a high dispersion of performance. For 50 universities there were no shifts in quartile.

*Table 5: Ranking lists of Italian universities by TS ratio and by $FSS_U$*

| University | Research staff | TS ratio | Rank | FSS | Rank | Rank shift | Percentile shift | Quartile shift |
|---|---|---|---|---|---|---|---|---|
| UNIV_64 | 68 | 48.5% | 1 | 3.081 | 1 | 0 | 0.0 | 0 |
| UNIV_6 | 61 | 36.1% | 2 | 1.763 | 2 | 0 | 0.0 | 0 |
| UNIV_8 | 57 | 29.8% | 3 | 1.659 | 4 | 1 | 1.6 | 0 |
| UNIV_7 | 39 | 28.2% | 4 | 1.735 | 3 | 1 | 1.6 | 0 |
| UNIV_45 | 239 | 19.7% | 5 | 1.257 | 6 | 1 | 1.7 | 0 |
| UNIV_34 | 1473 | 16.2% | 6 | 1.233 | 7 | 1 | 1.7 | 0 |
| UNIV_10 | 100 | 16.0% | 7 | 1.409 | 5 | 2 | 3.1 | 0 |
| UNIV_18 | 165 | 15.2% | 8 | 1.135 | 10 | 2 | 3.2 | 0 |
| UNIV_2 | 48 | 14.6% | 9 | 0.883 | 33 | 24 | 38.1 | 2 |
| UNIV_49 | 371 | 14.3% | 10 | 1.166 | 8 | 2 | 3.1 | 0 |
| UNIV_26 | 173 | 13.9% | 11 | 1.001 | 19 | 8 | 12.7 | 1 |
| UNIV_44 | 1177 | 13.8% | 12 | 1.084 | 12 | 0 | 0.0 | 0 |
| UNIV_29 | 1528 | 13.4% | 13 | 1.136 | 9 | 4 | 6.3 | 0 |
| UNIV_12 | 806 | 13.4% | 14 | 1.023 | 16 | 2 | 3.3 | 0 |
| UNIV_4 | 948 | 13.4% | 15 | 1.095 | 11 | 4 | 6.3 | 0 |
| UNIV_30 | 445 | 13.3% | 16 | 1.016 | 17 | 1 | 1.6 | 1 |
| UNIV_56 | 1695 | 13.2% | 17 | 1.032 | 14 | 3 | 4.7 | 1 |
| UNIV_55 | 276 | 13.0% | 18 | 0.919 | 26 | 8 | 12.7 | 0 |
| UNIV_51 | 164 | 12.8% | 19 | 0.967 | 20 | 1 | 1.6 | 0 |
| UNIV_22 | 412 | 12.6% | 20 | 0.909 | 28 | 8 | 12.7 | 0 |
| UNIV_59 | 450 | 12.4% | 21 | 0.942 | 23 | 2 | 3.3 | 0 |
| UNIV_15 | 270 | 11.9% | 22 | 1.035 | 13 | 9 | 14.2 | 1 |
| UNIV_5 | 652 | 11.7% | 23 | 0.893 | 31 | 8 | 12.8 | 0 |
| UNIV_16 | 114 | 11.4% | 24 | 0.935 | 24 | 0 | 0.1 | 0 |
| UNIV_60 | 1099 | 11.4% | 25 | 0.898 | 29 | 4 | 6.4 | 0 |
| UNIV_25 | 1216 | 11.3% | 26 | 0.950 | 21 | 5 | 7.9 | 0 |
| UNIV_14 | 205 | 11.2% | 27 | 0.942 | 22 | 5 | 7.9 | 0 |
| UNIV_62 | 429 | 10.7% | 28 | 0.934 | 25 | 3 | 4.8 | 0 |
| UNIV_54 | 480 | 10.6% | 29 | 1.006 | 18 | 11 | 17.4 | 0 |
| UNIV_61 | 464 | 10.3% | 30 | 1.029 | 15 | 15 | 23.7 | 1 |
| UNIV_37 | 689 | 10.3% | 31 | 0.909 | 27 | 4 | 6.3 | 0 |
| UNIV_53 | 278 | 10.1% | 32 | 0.835 | 35 | 3 | 4.8 | 1 |
| UNIV_46 | 439 | 10.0% | 33 | 0.725 | 48 | 15 | 23.9 | 0 |
| UNIV_38 | 785 | 9.3% | 34 | 0.897 | 30 | 4 | 6.3 | 1 |
| UNIV_40 | 1007 | 9.2% | 35 | 0.802 | 37 | 2 | 3.2 | 0 |
| UNIV_31 | 604 | 9.1% | 36 | 0.823 | 36 | 0 | 0.0 | 0 |

| University | Research staff | TS ratio | Rank | FSS | Rank | Rank shift | Percentile shift | Quartile shift |
|---|---|---|---|---|---|---|---|---|
| UNIV_42 | 521 | 9.0% | 37 | 0.796 | 38 | 1 | 1.7 | 0 |
| UNIV_27 | 922 | 8.9% | 38 | 0.761 | 43 | 5 | 8.0 | 0 |
| UNIV_47 | 417 | 8.4% | 39 | 0.847 | 34 | 5 | 7.9 | 0 |
| UNIV_32 | 1853 | 8.3% | 40 | 0.784 | 41 | 1 | 1.6 | 0 |
| UNIV_58 | 985 | 8.1% | 41 | 0.720 | 51 | 10 | 15.9 | 1 |
| UNIV_52 | 148 | 8.1% | 42 | 0.785 | 40 | 2 | 3.1 | 0 |
| UNIV_39 | 2642 | 7.8% | 43 | 0.740 | 47 | 4 | 6.4 | 0 |
| UNIV_9 | 660 | 7.7% | 44 | 0.752 | 44 | 0 | 0.0 | 0 |
| UNIV_24 | 119 | 7.6% | 45 | 0.789 | 39 | 6 | 9.4 | 0 |
| UNIV_20 | 990 | 7.2% | 46 | 0.748 | 46 | 0 | 0.1 | 0 |
| UNIV_36 | 710 | 7.0% | 47 | 0.723 | 50 | 3 | 4.8 | 1 |
| UNIV_35 | 1089 | 7.0% | 48 | 0.696 | 52 | 4 | 6.4 | 1 |
| UNIV_28 | 842 | 6.9% | 49 | 0.621 | 59 | 10 | 15.9 | 0 |
| UNIV_23 | 640 | 6.7% | 50 | 0.618 | 60 | 10 | 15.9 | 0 |
| UNIV_1 | 31 | 6.5% | 51 | 0.639 | 57 | 6 | 9.5 | 0 |
| UNIV_48 | 149 | 6.0% | 52 | 0.748 | 45 | 7 | 11.1 | 1 |
| UNIV_50 | 334 | 5.4% | 53 | 0.771 | 42 | 11 | 17.4 | 1 |
| UNIV_17 | 246 | 5.3% | 54 | 0.624 | 58 | 4 | 6.4 | 0 |
| UNIV_41 | 424 | 5.0% | 55 | 0.573 | 61 | 6 | 9.6 | 0 |
| UNIV_21 | 81 | 4.9% | 56 | 0.725 | 49 | 7 | 11.1 | 0 |
| UNIV_3 | 256 | 4.7% | 57 | 0.678 | 54 | 3 | 4.7 | 0 |
| UNIV_57 | 239 | 4.6% | 58 | 0.650 | 56 | 2 | 3.1 | 0 |
| UNIV_43 | 94 | 4.3% | 59 | 0.885 | 32 | 27 | 42.8 | 2 |
| UNIV_13 | 145 | 4.1% | 60 | 0.539 | 63 | 3 | 4.8 | 0 |
| UNIV_33 | 133 | 3.8% | 61 | 0.653 | 55 | 6 | 9.4 | 0 |
| UNIV_19 | 453 | 3.8% | 62 | 0.546 | 62 | 0 | 0.1 | 0 |
| UNIV_11 | 118 | 3.4% | 63 | 0.695 | 53 | 10 | 15.8 | 0 |
| UNIV_63 | 35 | 2.9% | 64 | 0.340 | 64 | 0 | 0.0 | 0 |

Table 6 presents the descriptive statistics of the quantile shifts and the correlation between rankings by TS ratio and by $FSS_U$ in each UDA. The last row shows values referring to all Italian universities without distinction per UDA. As expected, correlation between rankings is very strong in each UDA: the minimum Spearman coefficient of correlation (0.766) occurs in Chemistry (UDA 3) and the maximum (0.922) in Medicine (UDA 6). The largest percentage of universities shifting rank (97%) occurs in Earth Sciences (UDA 4) and Agricultural and veterinary sciences (UDA 7). Chemistry (UDA 3) registers the highest average percentile shift (15.4), while the maximum (61.9) occurs in Physics (UDA 2). In Chemistry, 50% of universities experience a quartile shift. The maximum quartile shift equals 2 for all UDAs with only one exception recorded for Medicine (UDA 6) where 12 out of 42 universities (or 29%) shift to a nearby quartile.

*Table 6: Descriptive statistics of quantile shifts and correlation between university rankings by TS ratio and by $FSS_U$*

| UDA* | N. of universities | Shifting in rank | Average percentile shift | Max percentile shift | Shifting quartile | Max quartile shift | Spearman correlation |
|---|---|---|---|---|---|---|---|
| 1 | 50 | 94% | 13.1 | 49.0 | 38% | 2 | 0.847 |
| 2 | 43 | 88% | 13.8 | 61.9 | 42% | 2 | 0.779 |
| 3 | 42 | 93% | 15.4 | 51.2 | 50% | 2 | 0.766 |
| 4 | 31 | 97% | 14.7 | 46.7 | 35% | 2 | 0.816 |
| 5 | 53 | 89% | 11.0 | 48.1 | 34% | 2 | 0.870 |
| 6 | 42 | 93% | 8.7 | 29.3 | 29% | 1 | 0.922 |
| 7 | 29 | 97% | 13.2 | 35.7 | 45% | 2 | 0.861 |
| 8 | 36 | 92% | 11.9 | 34.3 | 25% | 2 | 0.878 |
| 9 | 49 | 73% | 8.8 | 47.9 | 31% | 2 | 0.892 |
| All | 64 | 86% | 7.8 | 42.8 | 22% | 2 | 0.924 |

*\* 1 = Mathematics and computer science; 2 = Physics; 3 = Chemistry; 4 = Earth sciences; 5 = Biology; 6 = Medicine; 7 = Agricultural and veterinary sciences; 8 = Civil engineering; 9 = Industrial and information engineering*

The scatter plot of Figure 3 positions each university in terms of percentiles by TS ratio and $FSS_U$. Because top scientists have a very high effect on the overall rank by $FSS_U$ of a university (Abramo et al., 2013a), one expects that all universities fall either in the bottom-left quadrant or in the top-right one. The scatter plot helps to visualize anomalous occurrences. Three outliers (triangles) can be observed near the lines of the median values. Starting from the left side of the graph their position could be explained as follows. The position of UNIV_43 (7.9; 50.7) ranking just above the median by $FSS_U$ and among the bottom 10% by TS ratio, reveals that the productivity of its faculty is rather homogeneous. The opposite must be true for the UNIV_2 (87.3; 49.2), in the mirror position along the $FSS_U$ median: a high number of low and unproductive performers must offset the high contribution to the overall performance of a very high number of TSs. UNIV_46 (49.2; 25.3) positioned around the TS ratios median, must employ a very high number of professors whose productivity is far below the median.

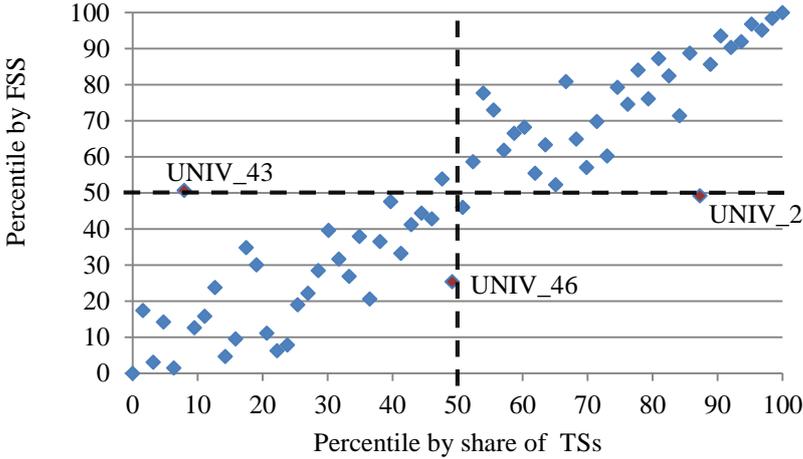

*Figure 3: Scatter plot of universities by their percentile ranks by $FSS_U$ and TS ratio*

## 5. Conclusions

In all productive sectors, competition is the lifeblood of continuous improvement and an unequalled stimulus in the search for excellence. Long-standing successful organizations are those that have developed the ability to attract and retain top players. The ratio of TSs to the overall faculty of a university can therefore be an indicator of the competitive strength of the university in the market for education. Up to 80% of world TSs in large research fields work in the top research institutions (Yang et al., 2015), which are concentrated in nations where competition among universities is strong. In non-competitive higher education systems, the research performance of universities is generally little differentiated. Because TSs contribute more than low performers in determining the position of universities in performance ranking lists (Abramo et al., 2013a), one expects that in non-competitive higher education systems the TSs will be rather evenly distributed among universities. In this study we confirm this expectation for the case of Italy. The differences in the ratios of TSs out of the overall faculty are negligible for all Italian universities, except for the three Schools for Advanced Studies and a private university focused in medicine, which outperform all others. Needless to say, these four universities are the ones that enjoy the highest reputation in the country in their respective educational programs, and are much sought out by prospective students.

As expected, the correlation between the ranking lists by ratio of TSs and average productivity is found to be very strong in all disciplines of the Sciences: universities which rank high by productivity are those which show also the higher TS ratios. Very few outliers escape this rule. One university in particular presents a very high ratio of TSs, but where their contribution to the average performance of the university is offset by a high ratio of low and unproductive performers. This instance is a paradox from a managerial standpoint and would prompt a case study to further investigate the issue. We have also found that the variability of TS ratios across SDSs and UDAs within single universities is very much higher than between the universities. These findings align with those referring to the variability of average productivity of overall academic staff within and between universities (Abramo et al., 2014).

The investigation also showed that there are no returns of TS ratio to size, which means that the size of universities does not favor or disfavor the emergence of TSs.

Future research on the topic might concern a comparison of the results for the Italian case with those for a country with a competitive higher education system.